\documentclass[nocompress]{spie}  

\usepackage{amsmath,amsfonts,amssymb}
\usepackage{graphicx}
\usepackage[colorlinks=true, allcolors=blue]{hyperref}
\usepackage{textpos}

\newcommand{\D}{{\rm{d}}}
\newcommand{\I}{{\rm{I}}}

\title{Characterization of free-space quantum channels }

\author[a, b]{D. Vasylyev}
\author[a, c]{A. A. Semenov}
\author[a]{W. Vogel}
\affil[a]{Institut f\"ur Physik, Universit\"at Rostock,
 Albert-Einstein-Str. 23, D-18059 Rostock, Germany}
\affil[b]{Bogolyubov Institute for Theoretical Physics, NAS of Ukraine, Vulytsya
	Metrologichna 14-b, 03680 Kiev, Ukraine}
\affil[c]{Institute of Physics, NAS of
Ukraine, Prospect Nauky 46, 03028 Kiev, Ukraine}
	
\authorinfo{Further author information: (Send correspondence to D.V.)\\D.V.: E-mail: dmytro.vasylyev@uni-rostock.de, Telephone: +49 381-498-6936}  

\begin{document}
\maketitle
\begin{textblock*}{7cm}(11cm,-5.2cm)
  \fbox {\footnotesize Proc. of SPIE \textbf{10771}, 107710V}
\end{textblock*}

\begin{abstract}
Many fundamental and applied experiments in quantum optics require transferring nonclassical states of light through large distances.
In this context the free-space channels are a very promising alternative to optical fibers as they are mobile and enable to establish communications with moving objects, using satellites for global quantum links.
For such channels the atmospheric turbulence is the main disturbing factor.
The statistical properties of the fluctuating transmittance through the turbulent atmosphere  are given by the probability distribution of transmittance (PDT).
We derive the consistent PDTs for the atmospheric quantum channels by step-by-step inclusion of various atmospheric effects such as beam wandering, beam broadening and deformation of the beam into elliptic form, beam deformations into arbitrary forms.
We discuss the applicability of PDT models for  different propagation distances and optical turbulence strengths in the case when the receiver module has an annular aperture.
We analyze the optimal detection and correction strategies which can improve the channel transmission characteristics.
The obtained results are important for the design of optical experiments including postselection and adaptive strategies and for the security analysis of quantum communication protocols in free-space.
\end{abstract}



\section{INTRODUCTION}
\label{sec:intro}

The experimental quantum atmospheric optics has grown quickly from  tabletop proof-of-principles experiments \cite{Bennett1989, Bennett1992} and has reached already the intercontinental scale by applying  satellites \cite{Liao2018}.
The interest to this subject rises from the attractive possibility of performing  quantum communication over atmospheric links \cite{Buttler1998, Hughes2002, Weier2006, Fedrizzi2009, Nauerth2013, Croal2016, Namazi2017, Yin2017}, to develop new measurement techniques \cite{Gottesman2012} , and to test the fundamental laws of physics \cite{Sabin2017, Handsteiner2017}.
The successful practical realization of free-space quantum protocols requires the development of advanced preselection, postselection, and adaptive strategies \cite{Vallone2015, Gruneisen2016, Bohmann2016, Wang2018}.

The main obstacle on the way of high-fidelity quantum communication in free-space is the atmospheric turbulence, random scattering, and absorption losses.
The absorption and scattering effects contribute merely to energy losses and the degradation of the signal intensity.
On the other hand, the atmospheric turbulence affects  amplitude and phase  of an optical beam in a random manner.
This leads to the degradation of spatial and temporal coherence of the light beam and causes scintillation, beam wandering, and phase front distortion.

Fluctuations of temperature, pressure, and humidity  of the air cause random variations of the atmospheric refractive index \cite{Tatarskii2016}.
The flow of turbulent air consists of a set of air blobs and vortices that spans a wide range of scales ranging from extremely large to very small.
Since the optical signal interacts along the propagation path with almost the  whole set of scales, the precise description of light propagation in turbulence is almost impossible and the free-space channel should be described by statistical means.
The statistical characteristics of the channel are then  related with the moments and correlation functions  of random optical amplitude and phase of the transmitted light.

The theory of classical light propagation through the atmospheric turbulence is well developed \cite{Tatarskii2016, Ishimaru1978, Andrews2001, Andrews2005}.
Some progress was also achieved in the theoretical description of free-space quantum light propagation \cite{Diament1970, Perina1973, Paterson2005} and in description of atmospheric quantum links \cite{Semenov2009, Semenov2010, Vasylyev2012, Vasylyev2016, Vasylyev2018}.
The atmosphere is considered as a linear lossy quantum channel characterized by fluctuating transmission properties.
The description of losses in quantum optics connects the annihilation operators of the input and output fields, $\hat a_{\mathrm{in}}$ and $\hat a_{\mathrm{out}}$, by the standard input-output relation
\begin{equation}
\label{eq:InputOutput}
\hat a_{\mathrm{out}}=\sqrt{\eta}\hat a_{\mathrm{in}}+\sqrt{1-\eta}\hat c \, .
\end{equation}
Here the operator $\hat c$ describes the environmental modes being in the vacuum state and $\eta\in[0,1]$ is the random channel transmittance.
The alternative representation of the input-output relation (\ref{eq:InputOutput}) can be given in terms of the Glauber-Sudarshan $P$ function~\cite{Glauber1963, Sudarshan1963}, which is a quasiprobability as it may attain negativities for nonclassical quantum states.
The relation between input $P_{\mathrm{in}}(\alpha)$ and output $P_{\mathrm{out}}(\alpha)$ states can be written as \cite{Semenov2009}
\begin{equation}
\label{eq:PInputOutput}
P_{\mathrm{out}}(\alpha)=\int\limits_0^1\D\eta\mathcal{P}(\eta)\frac{1}{\eta}P_{\mathrm{in}}\left(\frac{\alpha}{\sqrt{\eta}}\right) \, .
\end{equation}
Here $\mathcal{P}(\eta)$ is the probability distribution of the quantum channel transmittance (PDT) defined in the domain $\eta\in[0,1]$.
The knowledge of the PDT  suffices for the description of atmospheric quantum channel.

In the present article we review three theoretical models of atmospheric quantum channels.
The paper is organized as follows:
In Sec.~\ref{sec:Preliminaries} we briefly discuss various theoretical aspects needed for the description of light propagation in the turbulent atmosphere.
In Sec.~\ref{sec:bwm} we discuss the beam-wandering model for the calculation of the PDT.
In Sec.~\ref{sec:ElBeam} we consider the elliptic-beam PDT model. 
Using the law of total probability, we derive the general PDT in Sec.~\ref{sec:LawTotal}. 
Assuming  that the beam wandering effect is weak, we derive simple analytical expressions for the parameters of the corresponding conditional probability distribution.
Examples of PDTs for atmospheric links of various length are presented in Sec.~\ref{sec:Applications} and some techniques for improvements of the signal-to-nose ratio are discussed.
A summary is given in Sec.~\ref{sec:conclusions}.


\section{PRELIMINARIES}
\label{sec:Preliminaries}

In this paper, we review the theoretical models of the PDTs for the case  when the quantum light beams are prepared in the fundamental Gaussian mode.
While propagating in the atmospheric turbulence, a Gaussian beam undergoes random refraction and diffraction.
The detection of the transmitted beam will exhibit the random modulation of the received intensity.
Since the receiver has usually a finite-aperture collection/detection module, such as  a telescope  or a photodetector, this modulation coincides with the aperture transmittance of the incoming optical beam $\eta$.
The transmission properties of the aperture are closely related with the disturbances of the light beam during its propagation in the turbulent atmosphere.

Let us choose the coordinate system such that the beam propagates along $z$ axis onto the aperture plane at distance $z=L$.
 Then the intensity transmittance through the aperture reads as
\begin{equation}
\label{eq:Transm}
\eta =\int_{\mathcal{A}}\D^2\boldsymbol{r} I(\boldsymbol{r};L)\, .
\end{equation}
where $I(\boldsymbol{r};L)$ is the normalized intensity with respect to the full transversal plane $\boldsymbol{r}=\{x,y\}$ and $\mathcal{A}$ is the aperture area.

\begin{figure} [ht]
   \begin{center}
   \begin{tabular}{c}
   \includegraphics[height=6cm]{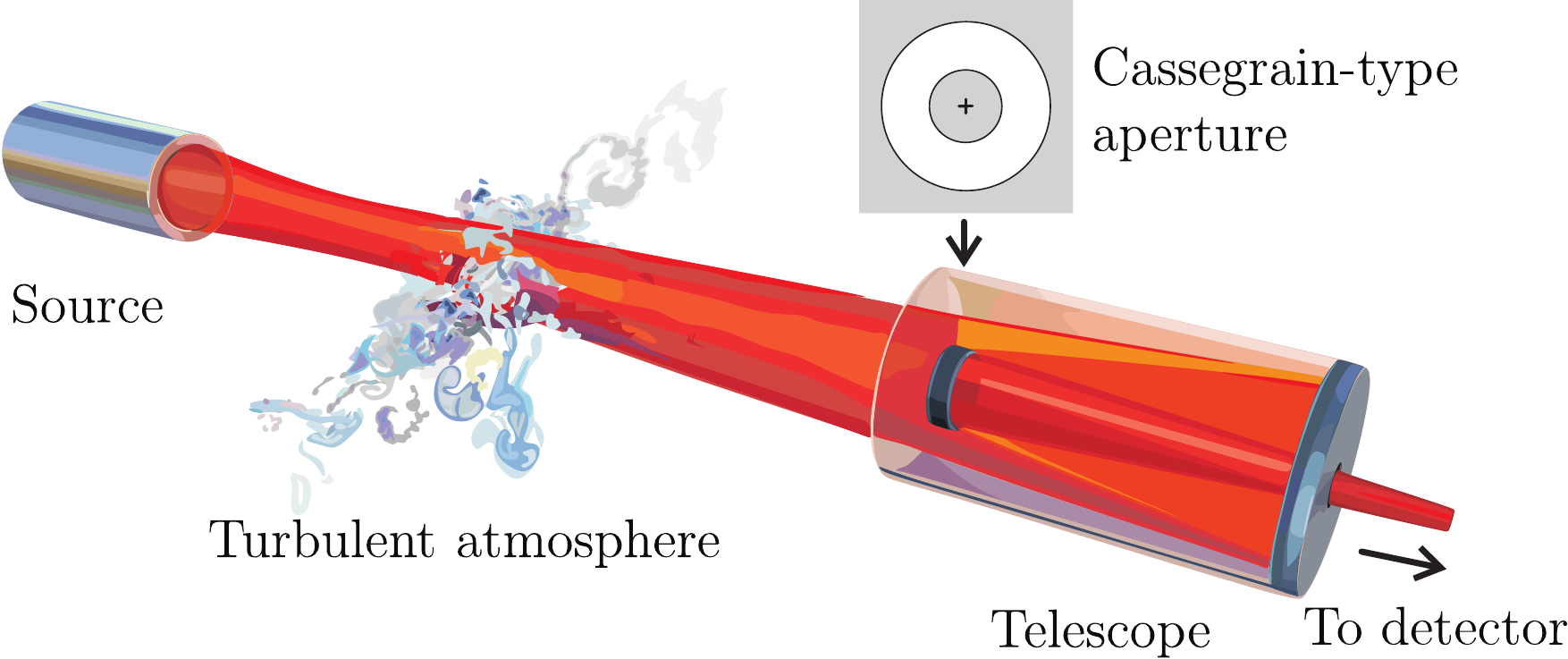}
	\end{tabular}
	\end{center}
   \caption[example]
   { \label{fig:Cassegr}
Optical communication scheme in free space. 
The optical beam generated by the transmitter source propagates through the atmospheric turbulence. 
The transmitted light  is then collected with the help of Cassegrain-type telescope  and is then  processed on the detection module.
The telescope aperture, which consists of large aperture opening of radius $a_1$ and the inner circular obscuration of radius $a_2$, is also shown. 
   }
   \end{figure}

In this article we consider the Cassegrain type aperture, i.e. the annular circular aperture with outer, $a_1$, and inner,  $a_2$, radii, cf.~Fig.~\ref{fig:Cassegr}.
This type of aperture is typical for Cassegrain reflecting telescopes which are the combination of two mirrors.
The smaller secondary mirror in this design appears as a circular obscuration that covers the center of the visible in focus larger aperture opening.
For this type of aperture the transmittance (\ref{eq:Transm}) can be written as
\begin{equation}
\label{eq:Transm1}
\eta = \eta_1 -\eta_2 =\left\{\int_{|\boldsymbol{r}|\le a_1} -\int_{|\boldsymbol{r}|\le a_2}\right\}\D^2\boldsymbol{r} I(\boldsymbol{r};L)\, .
\end{equation}
For definiteness in the following we use the values $a_1=0.075$~m and $a_2=0.023$~m for aperture radii, the values corresponding to $6$-inch reflector telescopes.

Using the definition (\ref{eq:Transm}) we can determine the important characteristics of atmospheric channels, namely the first two moments of channel transmittance
\begin{equation}
\label{eq:etaMean}
\langle\eta\rangle=\int\limits_{\mathcal{A}}\D^2\boldsymbol{r}\,\Gamma_2(\boldsymbol{r};L)\,,
\end{equation}
\begin{equation}
\label{eq:etaSquaredMean}
\langle\eta^2\rangle=\int\limits_{\mathcal{A}}\D^2\boldsymbol{r}_1\int\limits_{\mathcal{A}}\D^2\boldsymbol{r}_2\, \Gamma_4(\boldsymbol{r}_1,\boldsymbol{r}_2;L)\, .
\end{equation}
Here $\Gamma_2(\boldsymbol{r};L)=\langle I(\boldsymbol{r};L)\rangle$ and $\Gamma_4(\boldsymbol{r}_1,\boldsymbol{r}_2;L)=\langle I(\boldsymbol{r}_1;L)I(\boldsymbol{r}_2;L)\rangle$ are the second- and the fourth-order optical field correlation functions.
It appears\cite{Vasylyev2016, Vasylyev2018} that $\Gamma_2$ and $\Gamma_4$ play the central role in the description of atmospheric channels.
For example, these functions allow one to calculate the mean beam spot radius
\begin{equation}
\label{eq:Wst}
W_{\mathrm{ST}}=2\left[\int\limits_{\mathbb{R}^2}\D^2\boldsymbol{r}\, x^2\,\Gamma_2(\boldsymbol{r};L)- \sigma^2_{\mathrm{bw}} \right]^{1/2}\, ,
\end{equation}
where
\begin{equation}
\label{eq:BW}
\sigma^2_{\mathrm{bw}}=\langle x^2\rangle-\langle x\rangle^2=\int\limits_{\mathbb{R}^4}\D^2\boldsymbol{r}_1\D^2\boldsymbol{r}_2\,x_1\,x_2\,\Gamma_{4}(\boldsymbol{r}_1,\boldsymbol{r}_2;L)-\left(\int\limits_{\mathbb{R}^2}\D^2\boldsymbol{r}\, x\,\Gamma_2(\boldsymbol{r};L)\right)^2\,
\end{equation}
is the beam wandering variance, i.e.  the mean squared deviation of beam centroid position  relative to the aperture center.

The second and fourth coherence functions entering the expressions (\ref{eq:etaMean}) - (\ref{eq:BW}) can be calculated by using methods of classical atmospheric optics~\cite{Tatarskii2016}.
In many practically interesting situations the calculation of the $\Gamma_4$ function is connected with complicated numerical calculations~\cite{ Baskov2016, Baskov2018}
In this paper we use the phase approximation of the Huygens-Kirchhoff method \cite{Mironov1977, Aksenov1979} and  evaluate  $\Gamma_2$ and $\Gamma_4$ as
\begin{equation}
\label{eq:Gamma2PhAppr}
	\Gamma_2(\mathbf{r};L)=\frac{k^2}{4\pi^2L^2}\int_{\mathbb{R}^3}\D^2\mathbf{r}^\prime\,\exp\left[-\frac{|\mathbf{r}^\prime|^2}{2W_0^2}
	-2i\frac{\Omega}{W_0^2}\mathbf{r}{\cdot}\mathbf{r}^\prime{-}\frac{1}{2}\mathcal{D}_S(0,\mathbf{r}^\prime)\right]\, ,
\end{equation}
\begin{equation}
\label{eq:Gamma4PhAppr}
\begin{split}
		&\Gamma_4(\mathbf{r}_1,\mathbf{r}_2;L)=\frac{2k^4}{\pi^2(2\pi)^3L^4W_0^2}\int_{\mathbb{R}^6}\D^2\mathbf{r}_1^\prime
		\D^2\mathbf{r}_2^\prime\D^2\mathbf{r}_3^\prime e^{-\sum\limits_{i=1}^3\frac{|\mathbf{r}_i^\prime|^2}{W_0^2}-2i\frac{\Omega}{W_0^2}\left[(\mathbf{r}_1
		{-}\mathbf{r}_2){\cdot}\mathbf{r}_2^\prime{+}(\mathbf{r}_1{+}\mathbf{r}_2){\cdot}\mathbf{r}_3^\prime\right]}\\
		&\quad\times\exp\Bigl[\frac{1}{2}\sum\limits_{j=1,2}\Bigl\{\mathcal{D}_S(\mathbf{r}_1{-}\mathbf{r}_2,
		\mathbf{r}_1^\prime{+}(-1)^j\mathbf{r}_2^\prime)-\mathcal{D}_S(\mathbf{r}_1{-}\mathbf{r}_2,\mathbf{r}_1^\prime{+}(-1)^j\mathbf{r}_3^\prime)-
		\mathcal{D}_S(0,\mathbf{r}_2^\prime{+}(-1)^j\mathbf{r}_3^\prime)\Bigr\}\Bigr]\, ,
\end{split}
\end{equation}
where for the  Kolmogorov-Obukhov turbulence spectrum  \cite{Tatarskii2016} the phase structure function reads as
\begin{equation}
	\mathcal{D}_S(\mathbf{r},\mathbf{r}^\prime)=2C_n^2 k^2 L\int_0^1\D\xi\left|\mathbf{r}\xi+\mathbf{r}^\prime(1-\xi)\right|^{\frac{5}{3}}.
\end{equation}
	Here $k$ is the optical wavenumber,  $C_n^2$ is the turbulence refractive-index structure constant~\cite{Tatarskii2016}, $L$ is the propagation length, $W_0$ is the beam-spot width at the transmitter site, and $\Omega{=}kW_0^2/2L$ is the Fresnel parameter. 
	The so-called Rytov variance, $\sigma_R^2=1.23 C_n^2 k^{7/6} L^{11/6}$,  along with the parameter  $\Omega$ characterize the strength of optical turbulence~\cite{Charnotskii2012}.
	We adopt the simplified characterization of the optical turbulence strength, namely in the following we distinguish between weak ($\sigma_R^2<1$), moderate ($\sigma_R^2\approx1$) and strong ($\sigma_R^2\gg 1$) turbulence conditions.



\section{BEAM-WANDERING MODEL}
\label{sec:bwm}

In order to derive the simple PDT model that accounts only beam wandering effects, we derive firstly the transmission of a Gaussian laser beam through the Cassegrain-type aperture.
Let us assume that the beam has the fixed beam spot-radius $W$ at the aperture plane. 
Its value can be also taken to be equal to the mean beam spot-radius $W_{\mathrm{ST}}$ defined in  Eq.~(\ref{eq:Wst}).
The corresponding situation is depicted in Fig.~\ref{fig:Apert} a).
The intensity of the  beam with beam-spot radius $W$  focused at the aperture plane reads as
\begin{equation}
\label{eq:IntCirc}
I(\boldsymbol{r};L)=\frac{2}{\pi W^2}\exp\left[-\frac{2}{W^2}|\boldsymbol{r}-\boldsymbol{r}_0|^2\right] ,
\end{equation}
where $\boldsymbol{r}_0=(x_0\quad y_0)^{T}$ is the displacement vector of the beam centroid from the aperture center.
For simplicity we chose the coordinate system in such a way that the displacement $\boldsymbol{r}_0$ is aligned along the $x$ axis.
In this case, by substituting (\ref{eq:IntCirc}) in (\ref{eq:Transm}) and integrating in polar coordinates over the angle variable we arrive at
\begin{equation}
\label{eq:CircTransm}
\eta(r_0) =\frac{4}{W^2}\sum\limits_{n=1,2} (-1)^{n+1} e^{-2\frac{r_0^2}{W^2}}\int\limits_{0}^{a_n}\D r r e^{-2\frac{r^2}{W^2}}\I_0\left(\frac{4}{W^2}r_0 r\right) \, ,
\end{equation}
where $r_0=|\boldsymbol{r}_0|$, $\I_i(x)$ is the modified Bessel function of $i$th order.
The total transmittance through the annular aperture (\ref{eq:CircTransm}) simply  equals to the difference of transmittances through the apertures with radii $a_1$ and $a_2$.
The alternative representation of this result can be written also as the combination of two Marcum Q-functions~\cite{Marcum1950} 
\begin{equation}
\label{eq:CircTransm1}
\eta(r_0) =\sum\limits_{n=1,2} (-1)^{n}  Q\left(2\frac{r_0}{W},2\frac{a_n}{W}\right) \, ,
\end{equation}
where $Q(a,b)=\int_b^\infty\D x x\exp\left(-[x^2+a^2]/2\right)I_0(ax)$.
The integration in Eqs.~(\ref{eq:CircTransm}), (\ref{eq:CircTransm1}) should be performed numerically.
We have proposed~\cite{Vasylyev2012, Vasylyev2013} an accurate analytical approximations for such types of integrals.
Namely, we can rewrite  Eqs.~(\ref{eq:CircTransm}) and (\ref{eq:CircTransm1}) as
\begin{equation}
\label{eq:CircTransmApprox}
\eta(r_0) =\sum\limits_{n=1,2} (-1)^{n+1} \,\eta_0\left(a_n,2/W\right) \exp\left[-\left(\frac{r_0/a_n}{R(a_n,2/W)}\right)^{\lambda(a_n, 2/W)}\right] \, .
\end{equation}
Here
\begin{equation}
\label{eq:CircTransmEta0}
\eta_0(a_n,\xi) =1-\exp\left[-\frac{1}{2}a_n^2\xi^2\right] \,
\end{equation}
is the maximal transmittance through the aperture with radius $a_n$, i.e. the transmission under the condition $r_0=0$.
The scale, $R$, and shape, $\lambda$, parameters read as~\cite{Vasylyev2012}
\begin{equation}
\label{eq:CircTransmR}
R(a_n,\xi) =\left[\ln\left(\frac{2\eta_0(a_n,\xi)}{1-\exp[-a_n^2\xi^2]\I_0(a_n^2\xi^2)}\right)\right]^{-\frac{1}{\lambda(a_n,\xi)}} \, ,
\end{equation}
\begin{equation}
\label{eq:CircTransmlambda}
\lambda(a_n,\xi) =2 a_n^2\xi^2\frac{\exp[- a_n^2\xi^2]\I_1( a_n^2\xi^2)}{1-\exp[- a_n^2\xi^2]\I_0( a_n^2\xi^2)}\left[\ln\left(\frac{2\eta_0(a_n,\xi)}{1-\exp[- a_n^2\xi^2]\I_0( a_n^2\xi^2)}\right)\right]^{-1} \,,
\end{equation}
respectively. In the following we will also use the sort-hand notations $\eta_{0,n}$, $R_n$, and $\lambda_n$ for simplicity in notations.

For the derivation of the PDT that corresponds to the fluctuations of the transmittance (\ref{eq:CircTransmApprox}), we restrict our attention to the case when the main source of losses is the beam wandering.
In the case of isotropic turbulence we can assume that the beam deflection vector $\boldsymbol{r}_0$ is normally distributed with the variance $\sigma_{\mathrm{bw}}^2$ given by Eq.~(\ref{eq:BW}).
This assumption is equivalent to the Rayleigh probability density for the deflection parameter $r_0$
\begin{equation}
\label{eq:RayleighD}
\rho_{\mathrm{R}}(r_0;\sigma_{\mathrm{bw}}) =  \frac{r_0}{\sigma_{\mathrm{bw}}^2}\exp\left[-\frac{r_0^2}{2\sigma_{\mathrm{bw}}^2}\right]\,.
\end{equation}
The PDT can then be obtained by resolving the equation (\ref{eq:CircTransmApprox}) with respect to $r_0$ and performing the corresponding change of variables in the Rayleigh probability density.
This procedure is straightforward for a simple circular aperture\cite{Vasylyev2012}, but is quite cumbersome for the Cassegrain-type aperture and cannot be performed analytically.
In the latter case we can formally write for the PDT the following expression
\begin{equation}
\label{eq:PDTbw}
\mathcal{P}_{\mathrm{bw}}(\eta) =  \int\limits_{0}^\infty\D r_0 r_0 \rho_{\mathrm{R}}(r_0;\sigma_{\mathrm{bw}})\delta[\eta-\eta(r_0)]\,,
\end{equation}
where $\delta(x)$ is the Dirac delta function and $\eta(r_0)$ is given by Eq.~(\ref{eq:CircTransmApprox}).
The integration in Eq.~(\ref{eq:PDTbw}) should be performed numerically.

As a further generalization of the beam-wandering model we could further assume that the beam spot radius $W$ is also a random variable.
Indeed the light beam propagating in the turbulent atmosphere is randomly diffracted and this results in random broadening of the beam.
In the next section we go one step further and consider the most general situation when the beam remains Gaussian but it is broadened and deformed into elliptic form.



\section{ELLIPTIC-BEAM APPROXIMATION}
\label{sec:ElBeam}

We now extend the beam-wandering model described in the previous section and include the effects of beam broadening and deformation.
Here we restrict our attention to the specific set of deformations that preserve the Gaussian profile of the beam, i.e. we consider deformations of the Gaussian beam into the elliptic form\cite{Vasylyev2016}.
This approximation is appropriate for the regimes of weak and strong optical turbulence provided that the channel length is relatively short.

   \begin{figure} [ht]
   \begin{center}
   \begin{tabular}{c}
   \includegraphics[height=5cm]{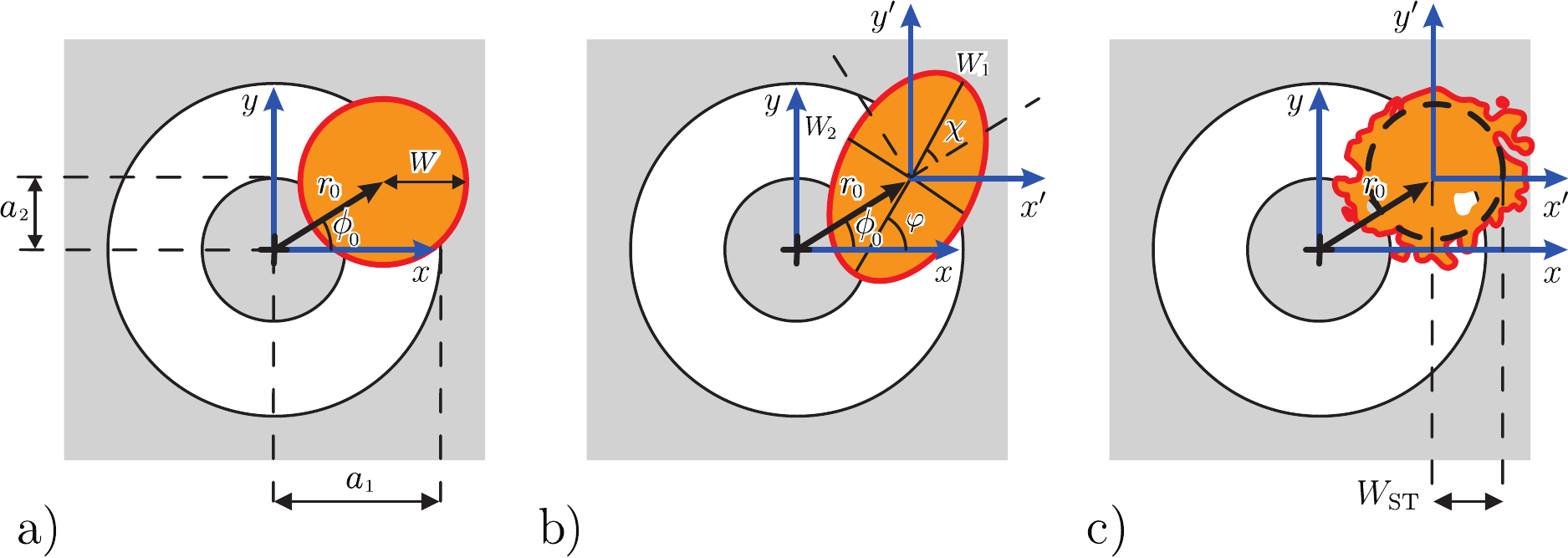}
	\end{tabular}
	\end{center}
   \caption[example]
   { \label{fig:Apert}
   Schematic representation  of beam spot distributions in the aperture plane are shown relative to aperture opening.
   The   Cassegrain-type annular aperture is given by the outer and inner  radii, $a_1$ and $a_2$, respectively.
   The centroid of the beam experiences the momentary displacement from the aperture center to the point $\boldsymbol{r}_0$ with the polar coordinates $(r_0, \phi_0)$.
   The $x^\prime-y^\prime$ coordinate system is associated with the beam centroid. 
   The diagrams represent the models considered in the article:
   a) The beam-wandering model.
   The beam is of circular form of radius $W$.
   b) The elliptic-beam model.
   The elliptical beam profile with the semi-axis $W_1$ rotated by the angle $\varphi$ relative to the $x$-axis and by the angle $\chi$ relative to the $\boldsymbol{r}_0$-associated axis is shown.
   c) More general situation described with the help of the law of total probability.
   The beam cross-section in this case has a complex structure and the beam dimensions are characterized by the mean short-term beam-spot radius $W_{\mathrm{ST}}$.
   }
   \end{figure}
   
In the case of weak turbulence, the propagating beam is not distorted too much by scattering on turbulent inhomogeneities and it preserves its initial Gaussian form.
If the turbulence  strength grows, the turbulent fluctuations distort the beam wave-front considerably.
This effect can be observed as a complex speckle pattern in the intensity distribution measured at the receiver cite.
For very strong turbulence, the intensity fluctuations are saturated, the speckle pattern is "washed out", and the beam profile is approximately  Gaussian.
Thus, in the regimes of weak and strong turbulence one can assume that the shape of the transmitted beam is formed by three major factors: beam wandering, beam broadening, and beam profile deformation into elliptic form.
The corresponding intensity distribution for both weak and strong turbulence can be written then as
\begin{equation}
\label{eq:IntEll}
I(\boldsymbol{r};L)=\frac{2}{\pi\sqrt{\det\boldsymbol{S}}}\exp\left[-2(\boldsymbol{r}-\boldsymbol{r}_0)^T\boldsymbol{S}^{-1}(\boldsymbol{r}-\boldsymbol{r}_0)\right].
\end{equation}
Here $\boldsymbol{S}$ is the real, symmetric, positive-definite spot-shape matrix.
The eigenvalues of this matrix, $W_i^2$, $i=1,2$, are squared semiaxes of the elliptic beam spot.
The semiaxis $W_1$ forms the angle $\varphi\in[0,\pi/2)$ relative to the $x$ axis, cf. Fig.~\ref{fig:Apert} b).
The set $\{W_1^2,W_2^2,\varphi\}$ uniquely describes the orientation and the size of the ellipse.
In the case of circular beam considered in the previous section,  it reduces to the diagonal matrix $\boldsymbol{S}=\mathrm{diag}\{W^2,W^2\}$ and Eq.~(\ref{eq:IntEll}) reduces to (\ref{eq:IntCirc}).
It is worth to note that a similar transmitted beam form was used~\cite{Baker2006} within the model of low-order turbulence phase fluctuations.
It has been shown~\cite{Baker2006} that this approximation is valid for atmospheric channels with weak turbulence and for short-distance atmospheric channels  with strong turbulence.

For elliptic-beam profile, the transmittance $\eta$ is obtained by substituting Eq.~(\ref{eq:IntEll}) into Eq.~(\ref{eq:Transm1}) and performing the corresponding integration over the Cassegrain aperture profile.
Unfortunately, the resulting integral cannot be evaluated analytically.
However, we can expect that the behavior of the transmittance as a function of the deflection vector $\boldsymbol{r}_0 = (r_0\cos\phi_0\quad r_0\sin\phi_0)^T$ cannot deviate significantly from those obtained previously for the circular aperture case.
Due to the elliptic form of the beam profile, the rotational symmetry is now broken.
Therefore, the transmittance depends also on the relative angle $\chi=\varphi-\phi_0$, cf. Fig.~\ref{fig:Apert} b).
We also observe that the transmittances $\eta_n$, $n=1,2$ in Eq.~(\ref{eq:Transm1})  behave similar to the transmittances of the circular Gaussian beams with the effective beam-spot radii
\begin{equation}
\label{eq:EffSpot}
W_{\mathrm{eff}}(\xi,a_n)=2 a_n\left[\mathcal{W}\left(\frac{4a_n^2}{W_1W_2}e^{\frac{a_n^2}{W_1^2}(1+2\cos^2\chi)+\frac{a_n^2}{W_2^2}(1+2\sin^2\chi)}\right)\right]^{-\frac{1}{2}}\, ,
\end{equation}
where $\mathcal{W}(x)$ is the Lambert $W$ function \cite{Corless1996}.
In this case the total transmittance $\eta$ can be approximated similarly to Eq.~(\ref{eq:CircTransmApprox}) and reads as
\begin{equation}
\label{eq:TransmEl}
\eta =\sum_{n=1,2}(-1)^{n+1}\eta_{0,n}\exp\left\{-\left[\frac{r_0/a_n}{R\left(a_n,\frac{2}{W_{\mathrm{eff}}(\chi,a_n)}\right)}\right]^{\lambda\left(a_n,\frac{2}{W_{\mathrm{eff}}(\chi,a_n)}\right)}\right\}.
\end{equation}
The  maximal possible transmittance $\eta_{0,n}$ for the aperture with radius $a_n$ can be written in terms of the incomplete Bessel function (or alternatively, the Marcum Q-function) and further approximated as
\begin{equation}
\label{eq:eta0Ellipt}
\begin{split}
\eta_{0,n} &=1-\I_0\left(a_n^2\left[\frac{1}{W_1^2}-\frac{1}{W_2^2}\right]\right)e^{-a_n^2[(1/W_1^2)+(1/W_2^2)]}\\
&-2\left[1-e^{-(a^2_n/2)[(1/W_1)-(1/W_2)]^2}\right]\exp\left\{-\left[\frac{\frac{(W_1+W_2)^2}{|W_1^2-W_2^2|}}{R\left(a_n,\frac{1}{W_1}-\frac{1}{W_2}\right)}\right]^{\lambda\left(a_n, \frac{1}{W_1}-\frac{1}{W_2}\right)}\right\}\, .
\end{split}
\end{equation}
It is worth to note that $\varphi$ is defined by modulo $\pi/2$ and hence the transmittance $\eta$ is a $\pi/2$-periodical function of $\varphi$~\cite{Vasylyev2016}.

The aperture transmittance (\ref{eq:TransmEl}) is a function of five real parameters, $\{x_0,y_0,\Theta_1,\Theta_2,\varphi\}$, randomly changed by the atmosphere, where $W_i^2=W_0^2\exp\Theta_i$ and $W_0$ is the initial beam-spot radius at the transmitter. We assume that the angle $\varphi$ is a $\pi/2$-periodical wrapped Gaussian variable~\cite{Mardia1999}.
With these assumptions, the PDT reads as
\begin{equation}
\label{eq:PDTelAp}
\mathcal{P}_{\mathrm{el}}(\eta) =  \frac{2}{\pi}\int\limits_{\mathbb{R}^4}\D^4 \boldsymbol{r} \int\limits_0^{\pi/2}\D\phi \rho_{\mathrm{G}}(\boldsymbol{V};\boldsymbol{\mu},\boldsymbol{\Sigma})\delta[\eta-\eta(\boldsymbol{V},\phi)]\,,
\end{equation}
where $\eta(\boldsymbol{V},\phi)$ is the transmittance defined in (\ref{eq:TransmEl}) and $\rho_{\mathrm{G}}(\boldsymbol{V};\boldsymbol{\mu},\boldsymbol{\Sigma})$ is the Gaussian probability density of the vector $\boldsymbol{V}$ with the mean $\boldsymbol{\mu}$ and the covariance matrix $\boldsymbol{\Sigma}$.
The vector of means $\boldsymbol{\mu}=(\langle x\rangle\quad\langle y\rangle\quad\langle\Theta_1\rangle\quad\langle\Theta_2\rangle)^T$ consists of
mean values of beam deflections
\begin{equation}
\langle x\rangle =\int\limits_{\mathbb{R}^2}\D^2\boldsymbol{r} x \Gamma_2(\boldsymbol{r};L),\qquad \langle y\rangle =\int\limits_{\mathbb{R}^2}\D^2\boldsymbol{r} y \Gamma_2(\boldsymbol{r};L)
\end{equation}
and the mean values of the log-variable $\Theta$
\begin{equation}
\label{eq:MeanTheta}
\langle\Theta_i\rangle=\ln\left[\frac{\langle W_i^2\rangle}{W_0^2}\left(1+\frac{\langle(\Delta W_i^2)^2\rangle}{\langle W_i^2\rangle^2}\right)^{-1/2}\right].
\end{equation}
Using the assumption of Gaussianity of variables and assuming that the turbulence is isotropic, we are able to simplify the covariance matrix  $\boldsymbol{\Sigma}$ considerably.
Its nonzero elements read as
\begin{equation}
\begin{split}
\label{eq:SigmaElements}
\Sigma_{11}&=\Sigma_{22}=\sigma_{\mathrm{bw}}^2,\\
\Sigma_{33}&=\Sigma_{44}=\langle(\Delta\Theta_1)^2\rangle=\ln\left[1+\frac{\langle(\Delta W_1^2)^2\rangle}{\langle W_1^2\rangle^2}\right],\\
\Sigma_{23}&=\Sigma_{32}=\langle\Delta\Theta_1\Delta\Theta_2\rangle=\ln\left[1+\frac{\langle\Delta W_1^2\Delta W_2^2\rangle}{\langle W_1^2\rangle \langle W_2^2\rangle}\right].
\end{split}
\end{equation}
Here the (co)variance elements $\langle \Delta W^2_i\Delta W^2_j\rangle$ are derived from the correlation functions
\begin{equation}
\begin{split}
\label{eq:WElements}
\langle W_i^2 W_j^2\rangle = 8\left\{-8\delta_{ij}(\sigma_{\mathrm{bw}}^2)^2 - \sigma_{\mathrm{bw}}^2\langle W_i^2\rangle +\int\limits_{\mathbb{R}^4}\D^2\boldsymbol{r}_1\D^2\boldsymbol{r}_2\left[x_1^2x_2^2(4\delta_{ij}-1)-x_1^2y_2^2(4\delta_{ij}-3)\right]\Gamma_4(\boldsymbol{r}_1,\boldsymbol{r}_2;L)\right\}
\end{split}
\end{equation}
and the mean squared beam-spot radii $\langle W_1^2\rangle$ and $\langle W_2^2\rangle$  are mutually equal and are given by Eq.~(\ref{eq:Wst}).

The elliptic-beam approximation is a reasonable approximation for short atmospheric quantum channels.
The PDTs obtained from Eq.~(\ref{eq:PDTbw}) show a good agreement with the experimental data for diverse daytime and weather conditions~\cite{Vasylyev2016, Vasylyev2017}.
However, for  longer   channels the moments of the distribution (\ref{eq:PDTbw}) start to deviate from the ones calculated from the first principles, i.e., from Eqs.~(\ref{eq:etaMean}) and (\ref{eq:etaSquaredMean}).
This makes the application of the elliptic-beam approximation doubtful for long-distance quantum channels.
In order to overcome the limitations of the elliptic-beam approximation, we have developed a new technique for the calculation of the PDT based on the law of total probability.



\section{LAW OF TOTAL PROBABILITY APPROACH AND WEAK BEAM WANDERING APPROXIMATION}
\label{sec:LawTotal}

The analysis of optical beam distortions caused by the atmospheric turbulence suggests that we can separate   beam wandering effects from the effects induced by  diffraction in the atmosphere.
Indeed, the refraction of the beam as a whole (beam wandering) is caused by turbulent inhomogeneities of sizes larger than the size of the beam cross-section.
On the other hand, the beam broadening, deformation and scintillation effects are caused by sizes of inhomogeneities smaller or comparable with the beam-spot diameter. 
This separation of scales and effects suggests that the PDT function can be written in the form where beam wandering is separated from beam broadening and deformation effects.
We have used~\cite{Vasylyev2018} the law of total probability \cite{Schervish} to derive the PDT based on these ideas.
Here we extend this model to more complex Cassegrain-type apertures. 

 We derive the PDT for  annular aperture from the joint probability density for beam transmittances $\eta_1$, $\eta_2$,  corresponding to the transmission through the apertures with radii $a_1$, $a_2$, respectively, cf.~ Fig.~\ref{fig:Apert} c).
Since the total transmittance $\eta$ is related to these partial transmittances as $\eta{=}\eta_1{-}\eta_2$,  the PDT for annular aperture can be written as
\begin{equation}
\label{eq:PDTCassAp}
\mathcal{P}(\eta)=\int\limits_0^1\D\eta_1\int\limits_0^{1-\eta}\D\eta_2\mathcal{P}_{1,2}(\eta_1, \eta_2)\delta(\eta-\eta_1+\eta_2)
= \int\limits_0^{1-\eta}\D\eta_2\mathcal{P}_{1,2}(\eta+\eta_2,\eta_2)\, ,
\end{equation}
where $\mathcal{P}_{1,2}(\eta_1, \eta_2)$ is the joint probability density for the partial transmittances $\eta_1$ and $\eta_2$.

We derive the joint probability $ \mathcal{P}_{1,2}(\eta_1,\eta_2)$  using the law of total probability on a similar footing as it was done for simple apertures~\cite{Vasylyev2018}.
The separation of beam wandering effects from those arising from beam broadening and deformation yields
\begin{equation}
\label{eq:Total}
 \mathcal{P}_{1,2}(\eta_1,\eta_2)=\int\limits_{\mathbb{R}^2}\D^2\boldsymbol{r}_0 P(\eta_1,\eta_2|\boldsymbol{r}_0)\rho (\boldsymbol{r}_0;\sigma_{\mathrm{bw}}).
\end{equation}
Here $P(\eta_1,\eta_2|\boldsymbol{r}_0)$ is the joint probability distribution of transmittance, conditioned on the beam centroid being displaced relative to the  centers of apertures by $\boldsymbol{r}_0$.
The  beam centroid displacement $\boldsymbol{r}_0$ is a random variable, the distribution of which is governed by the probability density $\rho (\boldsymbol{r}_0;\sigma_{\mathrm{bw}})$.
In the case of isotropic turbulence, the latter reduces to the Rayleigh distribution (\ref{eq:RayleighD}).

The analytical expression of the conditional probability $P(\eta_1,\eta_2|\boldsymbol{r}_0)$ can be deduced from the asymptotic analysis of  Eq.~(\ref{eq:Total}).
Beam wandering is a dominant effect for short distance atmospheric channels.
The corresponding asymptotic PDT is given by Eq.~(\ref{eq:PDTbw}). 
In the case when the beam wandering plays a minor role, such as for long distance atmospheric channels, the transmission statistics resembles the log-normal behavior \cite{Tatarskii2016, Vallone2015, Sun2013}.
Therefore, it is reasonable  to approximate the conditional probability for variables  $\boldsymbol{\eta}=(\eta_1\quad\eta_2)^T$ as the truncated bivariate log-normal distribution
\begin{equation}
\label{eq:CondProb}
 P(\eta_1,\eta_2|\boldsymbol{r}_0)=\left\{
 \begin{array}{l l}
 \frac{1}{\mathcal{F}}\frac{1}{2\pi\eta_1\eta_2\sqrt{\det\boldsymbol{\Sigma}}}\exp\left[-\frac{1}{2}(\ln\boldsymbol{\eta}-\boldsymbol{\mu})^T\boldsymbol{\Sigma}^{-1}(\ln\boldsymbol{\eta}-\boldsymbol{\mu})\right],& \eta_1\in[0,1],\quad \eta_2\in[0,1]\\
 0, &\text{else}
 \end{array}
\right. ,
\end{equation}
the truncation restricts the resulting PDT to the physical domain of the transmittance values. 
Here,  
\begin{equation}
 \mathcal{F}=\int\limits_0^1\D\eta_1\int\limits_0^1\D\eta_2\,\frac{1}{2\pi\eta_1\eta_2\sqrt{\det\boldsymbol{\Sigma}}}\exp\left[-\frac{1}{2}(\ln\boldsymbol{\eta}-\boldsymbol{\mu})^T\boldsymbol{\Sigma}^{-1}(\ln\boldsymbol{\eta}-\boldsymbol{\mu})\right]
\end{equation}
is the log-normal  cumulative distribution at the point $\eta_1=\eta_2=1$ used for normalization.
The parameters of the truncated bivariate log-normal distribution, $\boldsymbol{\mu}$ and $\boldsymbol{\Sigma}$, can be approximately replaced with the corresponding parameters of standard log-normal distribution, provided that $\mathcal{F}\approx 1$.
Using the subscripts $n,m=1,2$ for notation of quantities corresponding to apertures $a_1$ and $a_2$, we can approximately write
\begin{equation}
\label{eq:LNparameters}
 \mu_{n}(r_0) \approx -\ln\left[\frac{\langle\eta_n\rangle^2_{r_0}}{\sqrt{\langle\eta_n^2\rangle_{r_0}}}\right],\quad
\qquad \Sigma_{n,m}(r_0)\approx\sqrt{\ln\left[\frac{\langle\eta_n\eta_m\rangle_{r_0}}{\langle\eta_n\rangle_{r_0}\langle\eta_m\rangle_{r_0}}\right]},
\end{equation}
where the conditional  moments are related to the corresponding transmittance moments as
\begin{equation}
\label{eq:CondMoments}
 \langle\eta_n\rangle =\int\limits_{\mathbb{R}^2}\D^2\boldsymbol{r}_0\rho(\boldsymbol{r}_0;\sigma_{\mathrm{bw}})\langle\eta_n\rangle_{r_0},\qquad
 \langle\eta_n\eta_m\rangle =\int\limits_{\mathbb{R}^2}\D^2\boldsymbol{r}_0\rho(\boldsymbol{r}_0;\sigma_{\mathrm{bw}})\langle\eta_n\eta_m\rangle_{r_0}.
\end{equation}
Here the  correlation  $\langle\eta_n\eta_m\rangle$  is given by
\begin{equation}
 \langle\eta_n\eta_m\rangle = \int\limits_{|\boldsymbol{r}_1|\le a_n}\D^2\boldsymbol{r}_1  \int\limits_{|\boldsymbol{r}_2|\le a_m}\D^2\boldsymbol{r}_2 \Gamma_4(\boldsymbol{r}_1,\boldsymbol{r}_2;L)
\end{equation}
and the moments $\langle\eta_n\rangle$ and $\langle\eta_n^2\rangle$ are given by Eq.~(\ref{eq:etaMean}) and Eq.~(\ref{eq:etaSquaredMean}), respectively.

For determination of the parameters (\ref{eq:LNparameters}) and derivation of the conditional probability (\ref{eq:CondProb}) one has to resolve Eq.~(\ref{eq:CondMoments}) with respect to $\langle\eta\rangle_{r_0}$ and $\langle\eta^2\rangle_{r_0}$. 
This can be done by performing the inverse Weierstrass transform \cite{Brychkov1989}.
Alternatively one can apply here the approximative  approach  which is justified if the beam wandering effect is small~\cite{Vasylyev2018}.
Within the so-called weak beam wandering approximation we have the following analytical approximation formulas for the conditional moments:
\begin{equation}
\label{eq:etar0}
 \langle\eta_n\rangle_{r_0}=\eta_n^{(0)}\exp\left\{-\left[\frac{r_0/a_n}{R\left(a_n,\frac{2}{W_{\mathrm{ST}}}\right)}\right]^{\lambda\left(a_n,\frac{2}{W_{\mathrm{ST}}}\right)}\right\}\, ,
\end{equation}
\begin{equation}
\label{eq:etar20}
 \langle\eta_n\eta_m\rangle_{r_0}=\left[\zeta^{(0)}_{n,m}\right]^2\exp\left\{-\left[\frac{r_0/a_n}{R\left(a_n,\frac{2}{W_{\mathrm{ST}}}\right)}\right]^{\lambda\left(a_n,\frac{2}{W_{\mathrm{ST}}}\right)}-\left[\frac{r_0/a_m}{R\left(a_m,\frac{2}{W_{\mathrm{ST}}}\right)}\right]^{\lambda\left(a_m,\frac{2}{W_{\mathrm{ST}}}\right)}\right\}\, ,
\end{equation}
where the parameters $\eta_n^{(0)}$, $R_n$, and $\lambda_n$ are again given by Eqs.~(\ref{eq:CircTransmEta0}), (\ref{eq:CircTransmR}), and (\ref{eq:CircTransmlambda}), respectively.  It is worth to note that in the case when $\Gamma_2(\boldsymbol{r};L)$ significantly deviates from the Gaussian form, the prefactor $\eta_0$ in Eq.~(\ref{eq:etar0})  should be specified as
\begin{equation}
 \eta_n^{(0)}=\frac{\langle\eta_n\rangle}{\int\limits_0^\infty\D\xi \xi e^{-\frac{\xi^2}{2}}e^{-\left(\frac{\sigma_{\mathrm{bw}}}{R_n}\xi\right)^{\lambda_n}}}.
\end{equation}
This equation is derived via substituting Eq.~(\ref{eq:etar0}) in Eq.~(\ref{eq:CondMoments}), and then expressing $\eta_0$ explicitly.
In a similar way we determine the last unknown parameter $\zeta_{n,m}^{(0)}$ in (\ref{eq:etar20}),
\begin{equation}
\label{eq:zeta}
 \left[\zeta_{n,m}^{(0)}\right]^2=\frac{\langle\eta_n\eta_m\rangle}{\int\limits_0^\infty\D\xi \xi e^{-\frac{\xi^2}{2}}e^{-\left(\frac{\sigma_{\mathrm{bw}}}{R_n}\xi\right)^{\lambda_n}-\left(\frac{\sigma_{\mathrm{bw}}}{R_m}\xi\right)^{\lambda_m}}}.
\end{equation}
Therefore, the knowledge of seven parameters, $\langle\eta_1\rangle$, $\langle\eta_2\rangle$, $\langle\eta_1^2\rangle$, $\langle\eta_2^2\rangle$, $\langle\eta_1\eta_2\rangle$, $W_{\mathrm{ST}}$, and $\sigma_{\mathrm{bw}}^2$, which can be calculated from first principles, allows one to determine the conditional probability (\ref{eq:CondProb}) and hence the total PDT given by Eq.~(\ref{eq:Total}).

The weak beam wandering approximation is equivalent to the assumption that the conditional  intensity (co)variance  averaged over the aperture, $\langle\Delta\eta_n\Delta\eta_m\rangle_{r_0}$, does not depend on $\boldsymbol{r}_0$.
The experimental observations suggest that this approximation is true for long propagation channels \cite{Vorontsov2010, Gurvich2012}.
Inserting Eqs.~(\ref{eq:etar0}) and (\ref{eq:etar20}) into (\ref{eq:LNparameters}), we obtain explicit expressions for the parameters of the truncated log-normal distribution (\ref{eq:CondProb}),
\begin{equation}
\label{eq:weakBW}
 \mu_{n}(r_0)\approx -\ln\left[\frac{\left(\eta_n^{(0)}\right)^2}{\zeta^{(0)}_{n,n}}\right]+\left(\frac{r_0}{R_n}\right)^{\lambda_n},\qquad \Sigma_{n,m}(r_0)\approx\sqrt{\ln\left[\frac{\left(\zeta_{n,m}^{(0)}\right)^2}{\eta_n^{(0)}\eta_m^{(0)}}\right]}.
\end{equation}
As one can see, the weak beam wandering approximation yields the parameter $\boldsymbol{\Sigma}$ which does not depend on  beam deflection $r_0$.
This is not true for arbitrary atmospheric channels, i. e., in situations when the assumption of weak beam wandering is not valid.
The obtained parameters uniquely define the conditional PDT $P(\eta_1,\eta_2|\boldsymbol{r}_0)$, which is used in the law of total probability (\ref{eq:Total}) and in Eq.~(\ref{eq:PDTCassAp}) for determining the PDT of the channel under study.

We also note that the separation of beam wandering from the  other atmospheric disturbance effects in Eq.~(\ref{eq:CondProb}) allows us to describe the communication scenarios when partial or full beam tip tilt error correction is performed.
Usually such errors are mitigated by using the beam tracking procedure, which detects the position variations of a reference beam (beacon) sent by the receiver \cite{Tyler1994, Chun2017, Fernandez2018}.
The tracking procedure reduces the beam wandering variance to some new value $\Delta^2$.
By replacing $\sigma_{\mathrm{bw}}\rightarrow\Delta$ in Eqs.~(\ref{eq:CondProb})-(\ref{eq:zeta}) one  derives the PDT for quantum atmospheric link with the inclusion of the beam tracking procedure.



\section{APPLICATIONS}
\label{sec:Applications}

  We illustrate the proposed approaches for the PDT calculations by considering several atmospheric channels of diverse lengths. 
  The atmospheric refractive-index structure constant is chosen as $C_n^2=10^{-14}\,\text{m}^{-2/3}$, which corresponds to typical values for the atmospheric optical turbulence near the ground. 
  The wavelength of the optical field is chosen as $\lambda=800\, \text{nm}$ and the beam-spot radius at the transmitter source as $W_0=2$~cm. 
  
          \begin{figure} [ht]
   \begin{center}
   \begin{tabular}{c}
   \includegraphics[height=7cm]{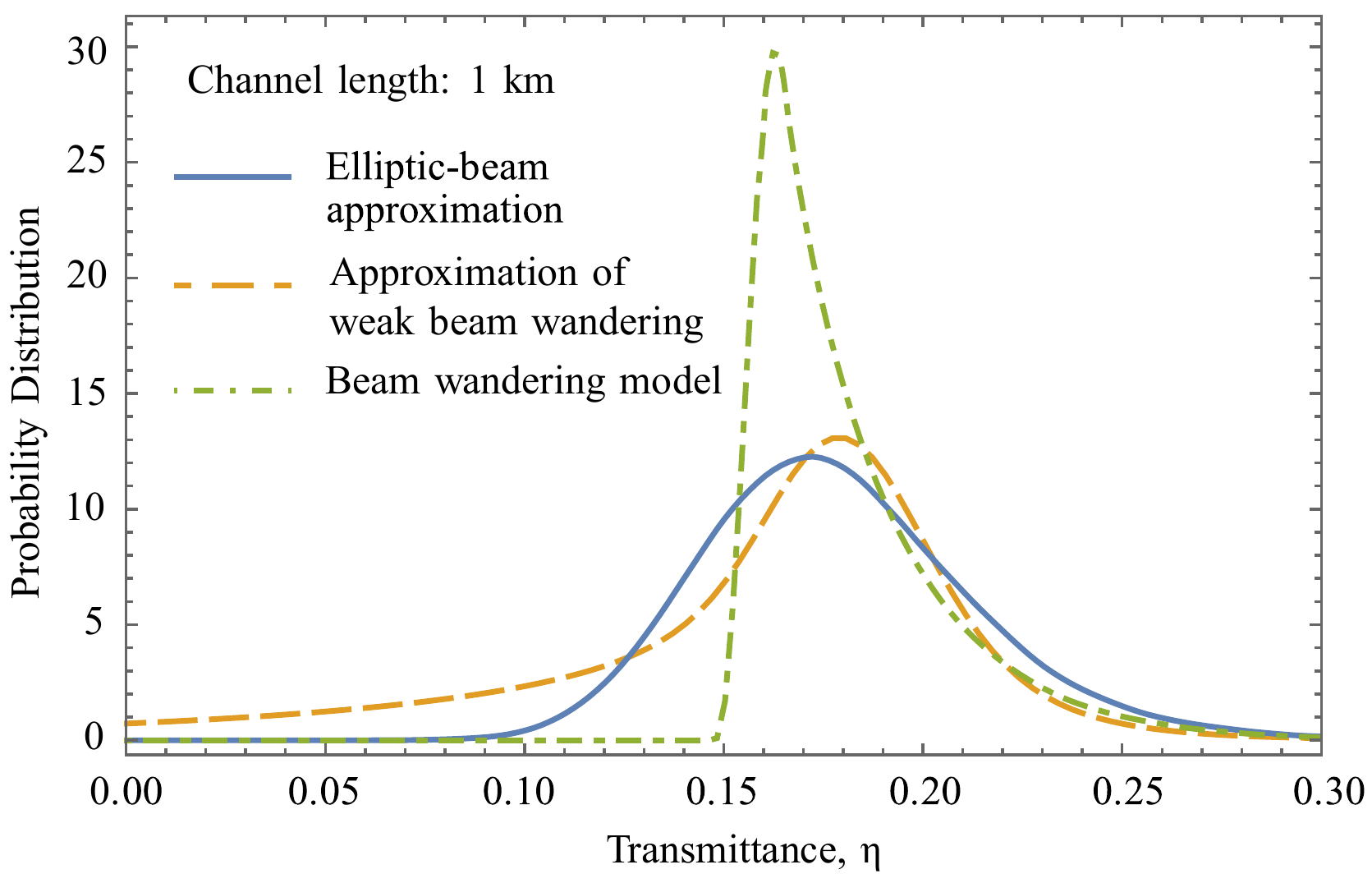}
	\end{tabular}
	\end{center}
   \caption[example]
   { \label{fig:3Methods}
  The  three PDT models considered in the article are compared for 1 km atmospheric link: the elliptic-beam approximation, the weak beam wandering approximation, and the beam-wandering model.
  The Rytov variance for this channel is $\sigma_R^2{=}0.43$ (weak optical turbulence).
  The additional deterministic attenuation of 2.3 dB due to molecular absorption~\cite{Elterman} is also included. 
   }
   \end{figure}

         \begin{figure} [ht]
   \begin{center}
   \begin{tabular}{c}
   \includegraphics[height=7cm]{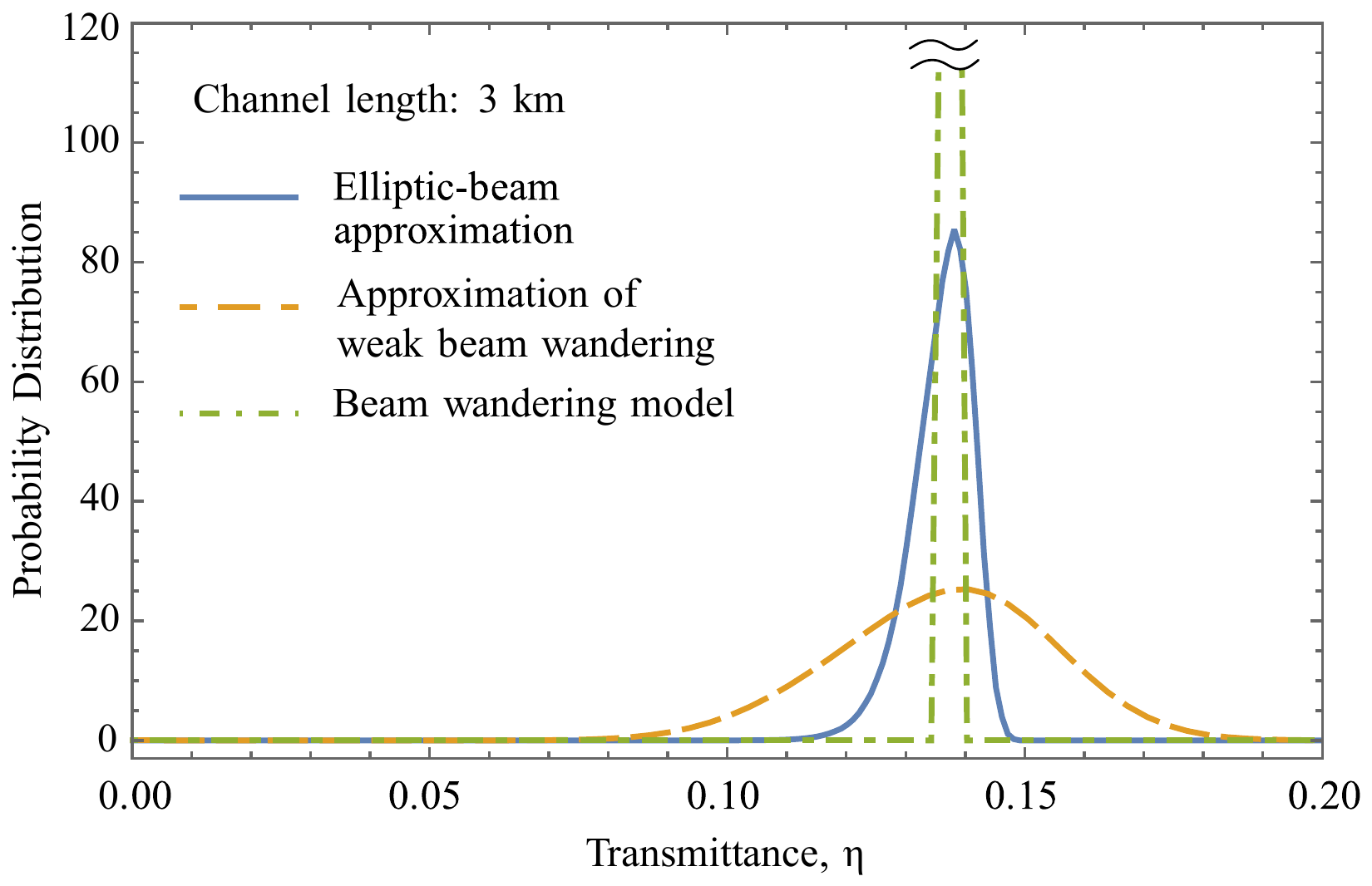}
	\end{tabular}
	\end{center}
   \caption[example]
   { \label{fig:3Methods3km}
Comparison of three PDT models for the  characterization of 3 km long atmospheric channel. 
The elliptic-beam approximation and the beam-wandering model yield incorrect values of the transmittance moments and hence do not describe the channel correctly.
   }
   \end{figure}
   
     \begin{figure} [ht]
   \begin{center}
   \begin{tabular}{c}
   \includegraphics[height=9cm]{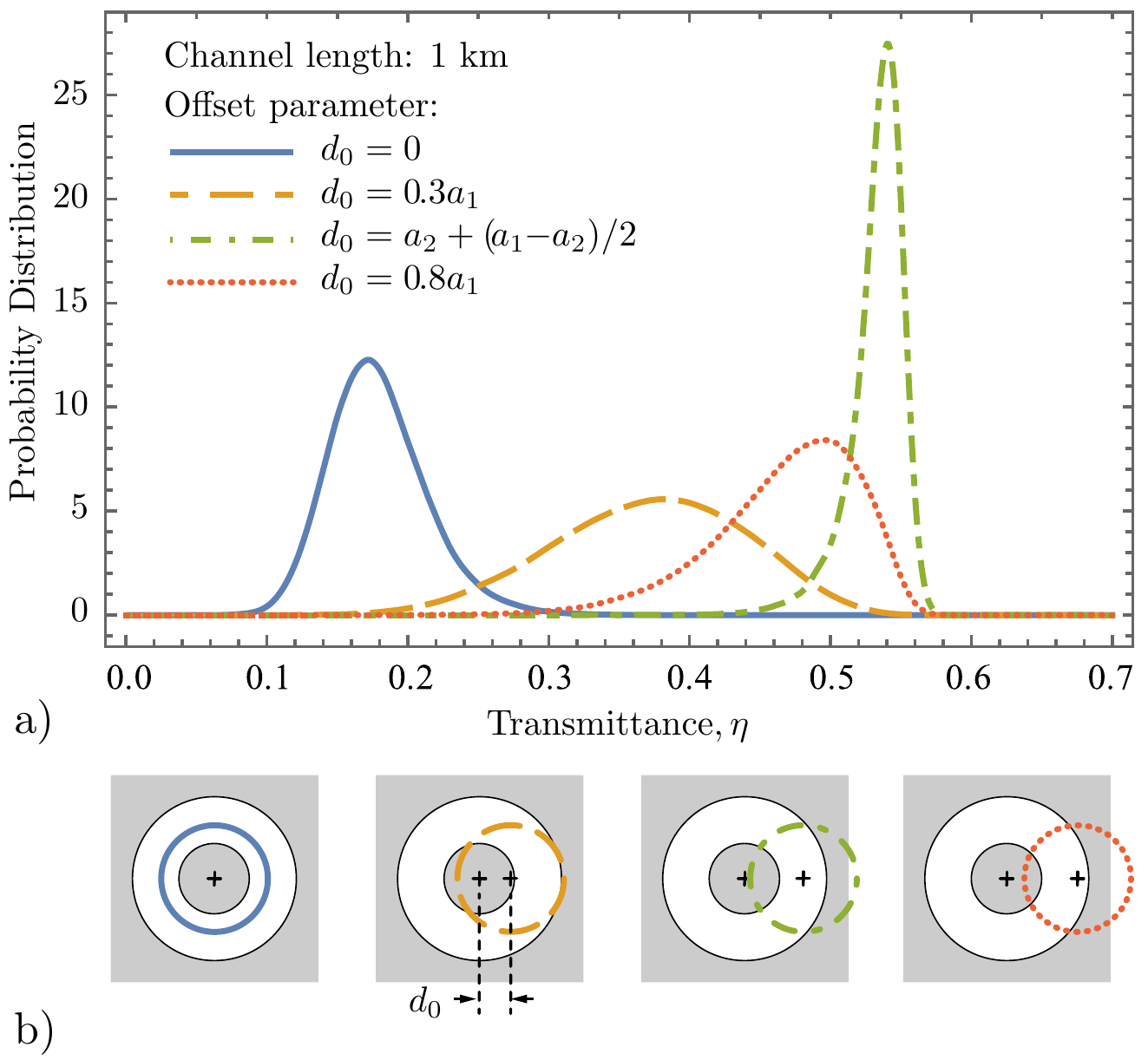}
	\end{tabular}
	\end{center}
   \caption[example]
   { \label{fig:Offset}
  The influence of beam centroid offset parameter $d_0$  on the form of the PDT for a 1 km channel a). 
  The schematic representation b) shows the corresponding offsets relative to the aperture center  together with the circles with radii $\sigma_\textbf{bw}$, cf. Eq.~(\ref{eq:BW}). 
  The aperture obscuration is shown by the gray area.
  The shift of the PDT to the larger values of the transmittance is governed by the decreasing overlap of beam transversal profile with the central aperture obscuration.
  After reaching the optimal value $d_0{=}a_2{+}(a_1{-}a_2)/2$, the influence of the outer aperture radius on beam truncation starts to grow and the aperture transmittance drops again.
   }
   \end{figure}  
   
      \begin{figure} [ht]
   \begin{center}
   \begin{tabular}{c}
   \includegraphics[height=7cm]{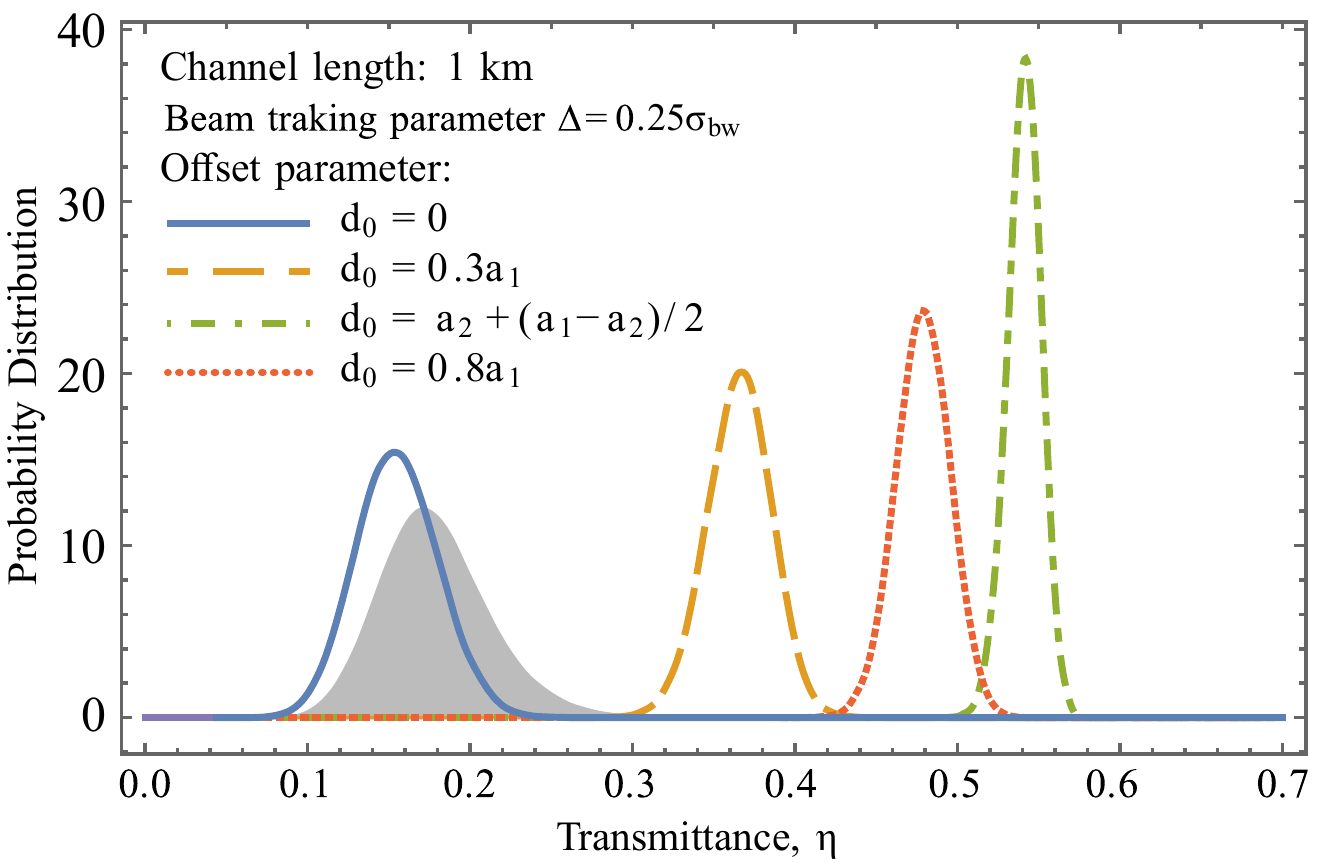}
	\end{tabular}
	\end{center}
   \caption[example]
   { \label{fig:OffsetTracking}
  The influence of the beam centroid offset parameter $d_0$ and the beam tracking on the form of the PDT for a 1 km channel.
  The tracking leads to a partial mitigation of beam wandering effects, such that the resulting beam wandering variance is given by $\Delta^2$. 
  As a reference, the PDT in the absence of beam tracking and with zero offset  is shown by the gray area, cf. also with the solid line in Fig.~\ref{fig:Offset}.
   }
   \end{figure}
   
  For short-distance  atmospheric channels the most appropriate PDT is obtained from the elliptic-beam approximation, cf.~Eq.~(\ref{eq:PDTelAp}).
  The corresponding PDT is shown in Fig.~\ref{fig:3Methods} as the solid line.
  The obtained PDT has moments which coincide with reasonable accuracy with those obtained from the first principles, i.e. from Eqs.~(\ref{eq:etaMean}), (\ref{eq:etaSquaredMean}).
  For the atmospheric channel under consideration the beam wandering effect contributes  significantly  to the total loss budget.
  As a consequence, the approximation of weak beam wandering based on Eq.~(\ref{eq:weakBW}) does not describe adequately the considered channel, cf.~Fig.~\ref{fig:3Methods}, dashed line.
  Finally, the PDT calculated within the beam wandering model (\ref{eq:PDTbw})  shows a sharp peak and a discontinuity-like behavior for small transmittances, cf.~Fig.~\ref{fig:3Methods}, dash-dotted line.
  This behavior is expectable since the beam-wandering model does not account for random beam broadening and deformation effects, which usually smear out the sharp edges of the  beam-wandering PDT.

  In the case of long propagation channels the beam-wandering model shows narrow peaks centered near the mean transmittance values.
  This behavior is clearly seen in Fig.~\ref{fig:3Methods3km} (dash-dotted line) for a 3 km atmospheric link.
  The small width of this distribution suggests that the beam wandering is relatively small and that the weak beam wandering approximation considered in Sec.~\ref{sec:LawTotal} is valid for the description of the channel, cf. dashed line in Fig.~\ref{fig:3Methods3km}.
  The elliptic-beam approximation (\ref{eq:PDTelAp}) yields incorrect values for the mean transmittance and its variance.
  That is, it is not appropriate for modeling  atmospheric channels of such lengths, cf. solid line in  Fig.~\ref{fig:3Methods3km}.

  In Figure~\ref{fig:Offset} we illustrate the influence of inner obscuration of the receiver Cassegrain-type telescope on the transmittance characteristics of short distance channel.
  The corresponding calculations are based on the elliptic-beam approximation for the PDT.
  In the case when the mean beam wandering displacement, $\langle \boldsymbol{r}_0\rangle$, is zero, i.e. when the center of beam wandering fluctuations coincides with the aperture center, the aperture inner circular obscuration with radius $a_2$  truncates most of the times the central part of the transmitted beam.
  This leads to the considerable diminishing of total channel transmittance.
  By increasing the so-called offset parameter $d_0=\langle|\boldsymbol{r}_0|\rangle$ we can improve the transmittance. 
  The optimal value for the offset parameter for Cassegrain-type aperture is given roughly by  $d_{\mathrm{opt}}\approx d_0=a_2+(a_1-a_2)/2=(a_1+a_2)/2$, i.e. when it coincides with the mean value of two aperture radii. 
  This situation corresponds to the case when the beam is targeted not on the aperture center but on the aperture annular opening half-way the distance $a_1{-}a_2$ from the edge of the inner obscuration.
  The use of this optimal beam offset for short propagation distances or small beam spot sizes  leads to a considerable  increase of the channel transmittance and improves the signal-to-noise ratio as clearly seen in Fig.~\ref{fig:Offset} by comparing the solid line and the dash-dotted. 
  
  Figure \ref{fig:OffsetTracking} shows the PDTs for a 1 km link for various offset parameters and with the partial compensation of beam wandering.
  The beam wandering variance after applying the beam tracking procedure is reduced to $\Delta^2 =0.25^2\sigma_{\mathrm{bw}}^2$.
  By comparing curves in Fig.~\ref{fig:OffsetTracking} for the scenario of beam tracking and non-zero offset parameter with the corresponding curves without beam tracking in Fig.~\ref{fig:Offset}, we can see that the beam tracking procedure improves the channel transmittance  only for the optimal value $d_{\mathrm{opt}}$ of the offset parameter.
  In other cases the PDTs become narrower with the consequence of reducing the maximally possible values of the channel transmittance.
  Therefore, for quantum protocols that utilize the postselection strategies\cite{Vallone2015, Wang2018, Gumberidze2016} based on the selection of transmission events with the transmittance above some threshold values, the additional beam tracking does not always improve the protocol performance.
  The best strategy in such cases is the setting of the offset parameter to its optimal value.
  Additionally, for long propagation channels we have not found any significant improvement by using beam tracking and non-zero offset. 
  We note that the discussed results are  applicable to annual-type apertures.

   \begin{figure} [ht]
   \begin{center}
   \begin{tabular}{c}
   \includegraphics[height=7cm]{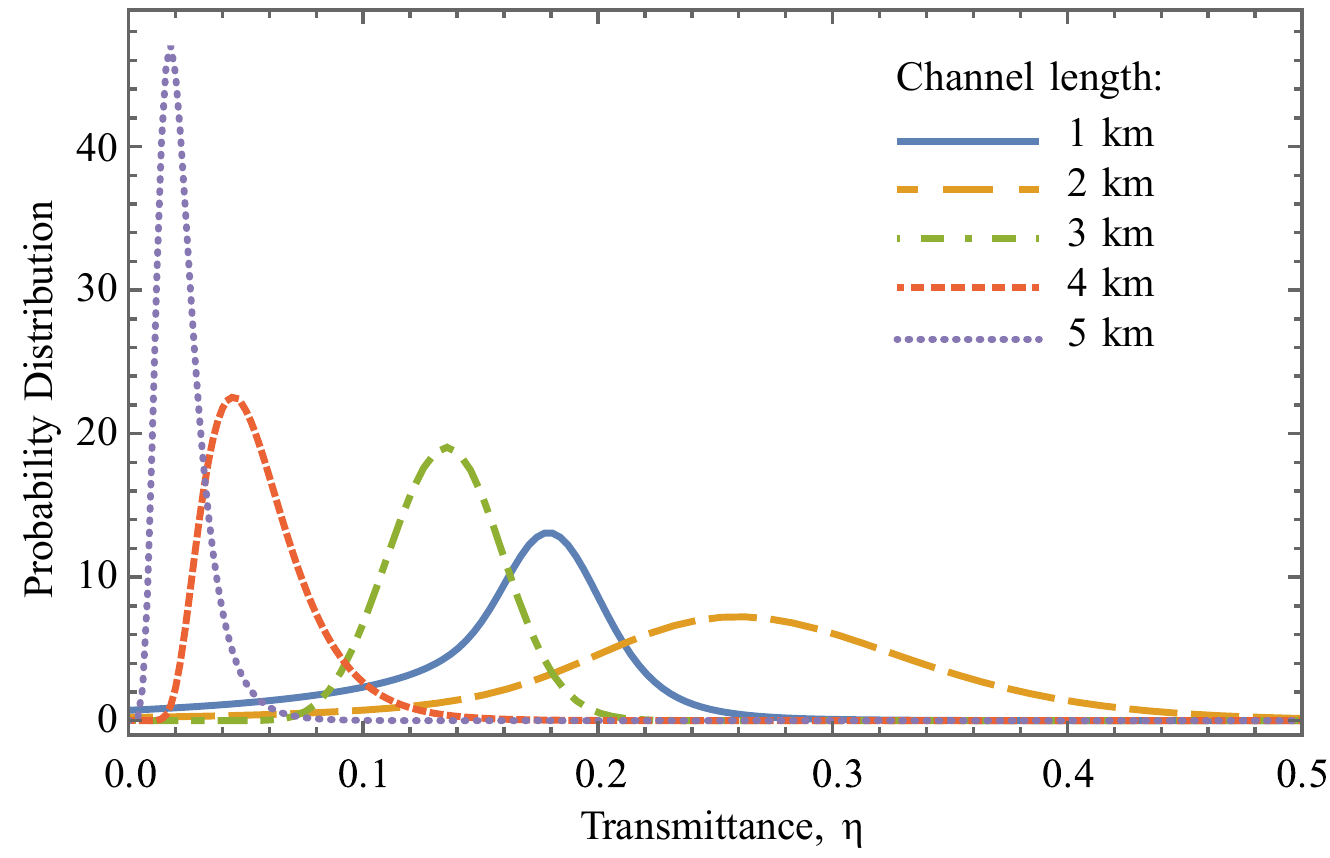}
	\end{tabular}
	\end{center}
   \caption[example]
   { \label{fig:LawTotal}
  The PDTs calculated within the weak beam wandering approximation for atmospheric channels of various lengths.
  The Rytov variances for 1 km, 2 km, 3 km, 4 km, and 5 km propagation length are $\sigma_R^2 =0.43$, $1.53$,  $3.22$, $5.47$, and $8.23$, respectively.
  Thus, the plot illustrates the transition from weak to strong optical turbulence.
  The worse transmission performance for a 1 km link in comparison with a 2 km channel is explained by almost complete  shadowing of the transmitted beam by the central obscuration of the receiver aperture.
  The corresponding additional deterministic losses~\cite{Elterman} are also included.
   }
   \end{figure}
   
  Figure \ref{fig:LawTotal} shows the PDTs for various propagation lengths calculated using the weak beam wandering approximation.
  With  increasing propagation length,  the influence of turbulent distortion factors on the propagating beam grows as well.
  Since the beam wandering effect shows saturation in the strong turbulence regime~\cite{Mironov1976}, for long distance channels the effects of beam broadening and deformation play a crucial role.
  In comparison with the PDT of a 2 km optical link, the PDT of a 1 km channel is shifted towards  smaller values of the transmittance.
  This is a direct consequence of beam truncation by the central aperture obscuration as was discussed above.
  This result agrees with the observation that the annular aperture degrades the signal-to-noise ratio in astronomical photometry~\cite{Young1967}.
  By choosing the appropriate offset parameter, one can improve the channel transmission characteristics (cf.~Fig.~\ref{fig:Offset}).
  However this strategy works for transmitted  beams with beam spot radii smaller or comparable with the radius of the inner obscuration.
  For wider beams or for longer propagation lengths the introduction of an offset would  lead to increasing average losses.

\section{CONCLUSIONS}\label{sec:conclusions}

Free-space quantum channels are characterized by the  probability distribution of the random transmittance.
The intensity of transmittance fluctuations is dictated by the receiver aperture characteristics, while the statistical properties of transmittance are related with the atmosphere-induced  effects such as beam wandering, beam broadening, and beam deformation.  
We reviewed three models of the probability distribution of transmittance based on the stepwise inclusion of these atmospheric effects.
The situation when the receiver uses the Cassegrain-type telescope for the collection of the transmitted signal has been considered.
The corresponding  aperture has an annular form.

The first model of quantum channel considers the situation when the turbulence induces beam wandering effects only.
In this model, the Gaussian beam is randomly reflected by  turbulent inhomogeneities which cause random wandering of beam centroid relative to the aperture center.
The shape of the beam remains circular.
For a simple circular aperture the corresponding probability distribution for transmittance of the wandering beam can be derived analytically and is the log-negative Weibull distribution.
In the case of a Cassegrain-type aperture the analytical presentation is not possible and the probability distribution has been  evaluated here by numerical means.

The refraction of an optical beam is possible when the size of turbulent inhomogeneities is larger than the beam cross-section diameter.
However, the atmospheric turbulence  is a rather complex composition of inhomogeneities of different scales and shapes.
The scales of the order of the beam diameter contribute to the beam diffraction-induced broadening and deformation.
These additional effects are taken into account by the so-called elliptic-beam model and the model based on the law of total probability.
The elliptic-beam  model considers beam deformation into elliptic forms and is applicable for short propagation distances.
The model based on the law of total probability is applicable for transmitted beams with more complex deformations.
It allows one to separate the contributions from beam wandering and beam broadening/deformation  and hence to describe beam tracking procedures that are often used for the tip-tilt compensation.
With the assumption of weak beam wandering the corresponding probability distribution of transmittance can be derived in a simple form. 
Such simplified model is reasonable for the description of long-distance atmospheric channels.
In the present article we derived the corresponding probability distributions for the case when the receiver collects the transmitted light with the Cassegrain-type aperture.
We summarize the regions of applicability of each model considered here in Table~\ref{tab:Models}.
 
   \begin{table}[ht]
\caption{Applicability range of the considered models for the PDT and the corresponding included atmospheric disturbances.} 
\label{tab:Models}
\begin{center}       
\begin{tabular}{|l|l|l|l|}
\hline
\rule[-1ex]{0pt}{3.5ex}  Model & Included effects & Applicability range & Rytov parameter\\
\hline
\hline
\rule[-1ex]{0pt}{3.5ex}  Beam-wandering model & beam wandering & short distance channel &$\sigma_R^2 >0$\\
\hline
\rule[-1ex]{0pt}{3.5ex}  Elliptic-beam model & beam wandering, beam broadening& short distance channel &$  \sigma_R^2> 0$ \\
\rule[-1ex]{0pt}{3.5ex}  & beam deformation into elliptic form &$L < 2$ km& \\
\hline
\rule[-1ex]{0pt}{3.5ex}  Approximation of weak &  beam wandering, beam broadening & long distance channel&   $\sigma_R^2 >1$ \\
\rule[-1ex]{0pt}{3.5ex}  beam wandering & beam deformation into arbitrary form & $L>2$ km & \\
\hline 
\end{tabular}
\end{center}
\end{table}

The analysis of the PDTs for short propagation distances or small beam-spot radii of the transmitted light has shown that the transmission statistics is greatly influenced by the central obscuration of the receiver Cassegrain-type aperture.
In order to correct the destructive effects caused by the central obscuration we propose to use nonzero offset between the aperture center and the center of beam wandering fluctuations of the transmitted beam.
The correction of beam-wandering fluctuations by applying tracking procedures could improve the channel transmittance only when the optimal offset is used.
For longer atmospheric channels the introduction of an additional offset or beam tracking will not improve the channel transmission characteristics considerably.
This findings are important for designing adaptive techniques in  experiments with quantum light which involve  receiver modules with Cassegrain-type telescopes.
In the companion paper we  will apply the PDT models derived here for the analysis  of  quantum properties of the transmitted nonclassical light fields.

\acknowledgments D. V. is grateful to A. Korzhuev for useful and enlightening discussions.
The work was supported by the Deutsche Forschungsgemeinschaft through
projects VO 501/21-2.

\end{document}